\documentclass[twoside]{article}

\usepackage[accepted]{aistats2022}
\usepackage[round]{natbib}

\usepackage{graphicx}
\usepackage{subcaption}
\usepackage{hyperref}
\hypersetup{
colorlinks=true,
linkcolor=red,
citecolor=blue,
}
\usepackage{xcolor}
\usepackage{amsfonts,amsmath,amssymb}
\usepackage{url}
\usepackage{booktabs}
\usepackage{makecell}
\usepackage{float}
\newcommand{\ie}{\em{i.e.}}
\newcommand{\eg}{\em{e.g.}}
\usepackage{cleveref}
\usepackage{multicol,multirow}
\usepackage{enumitem }
\usepackage{bm}

\def\eqref#1{equation~\ref{#1}}
\def\1{\bm{1}}

\def\ve{{\bm{e}}}

\def\vh{{\bm{h}}}

\def\vx{{\bm{x}}}
\def\vy{{\bm{y}}}
\def\vz{{\bm{z}}}

\def\mE{{\bm{E}}}

\DeclareMathAlphabet{\mathsfit}{\encodingdefault}{\sfdefault}{m}{sl}
\SetMathAlphabet{\mathsfit}{bold}{\encodingdefault}{\sfdefault}{bx}{n}

\def\gG{{\mathcal{G}}}

\def\gL{{\mathcal{L}}}

\def\sN{{\mathbb{N}}}

\def\sR{{\mathbb{R}}}

\usepackage{cancel}
\usepackage{adjustbox}

\begin{document}

% If your paper is accepted and the title of your paper is very long,
% the style will print as headings an error message. Use the following
% command to supply a shorter title of your paper so that it can be
% used as headings.
%
%\runningtitle{I use this title instead because the last one was very long}

% If your paper is accepted and the number of authors is large, the
% style will print as headings an error message. Use the following
% command to supply a shorter version of the authors names so that
% they can be used as headings (for example, use only the surnames)
%
%\runningauthor{Surname 1, Surname 2, Surname 3, ...., Surname n}

\runningauthor{Shengchao Liu, Meng Qu, Zuobai Zhang, Huiyu Cai, Jian Tang}

\twocolumn[

\aistatstitle{Structured Multi-task Learning for Molecular Property Prediction}

\aistatsauthor{
Shengchao Liu\textsuperscript{1,2},~~~
Meng Qu\textsuperscript{1,2},~~~
Zuobai Zhang\textsuperscript{1,2},~~~
Huiyu Cai\textsuperscript{1,2},~~~
Jian Tang\textsuperscript{1,3,4}
}

\aistatsaddress{
\textsuperscript{1}Mila~~~~
\textsuperscript{2}Université de Montréal~~~~
\textsuperscript{3}HEC Montréal~~~~
\textsuperscript{4}CIFAR AI Chair
}

]

\begin{abstract}
Multi-task learning for molecular property prediction is becoming increasingly important in drug discovery. However, in contrast to other domains, the performance of multi-task learning in drug discovery is still not satisfying as the number of labeled data for each task is too limited, which calls for additional data to complement the data scarcity. In this paper, we study multi-task learning for molecular property prediction in a novel setting, where a relation graph between tasks is available. We first construct a dataset (ChEMBL-STRING) including around 400 tasks as well as a task relation graph. Then to better utilize such relation graph, we propose a method called SGNN-EBM to systematically investigate the structured task modeling from two perspectives. (1) In the \emph{latent} space, we model the task representations by applying a state graph neural network (SGNN) on the relation graph. (2) In the \emph{output} space, we employ structured prediction with the energy-based model (EBM), which can be efficiently trained through noise-contrastive estimation (NCE) approach. Empirical results justify the effectiveness of SGNN-EBM. Code is available on \href{https://github.com/chao1224/SGNN-EBM}{the GitHub repository}.
\end{abstract}

\section{Introduction} \label{sec:intro}
Predicting the properties of molecules ({\eg}, binding affinity with proteins, toxicity, ADME property) is a fundamental problem in drug discovery. Recently, we witness many successes of deep neural networks for molecular property prediction~\cite{dahl2014multi,unterthiner2014deep,ramsundar2015massively,ramsundar2017multitask,wu2018moleculenet,liu2018practical,liu2019loss,hu2019strategies,rong2020self,alnammi2021evaluating,liu2021pre}. In particular, molecules are represented as molecular graphs, and graph neural networks~\cite{kipf2016semi}---which are neural network architectures specifically designed for graphs---are utilized for learning molecular representations. These neural networks are then usually trained with a set of labeled molecules. However, one big limitation for property prediction in drug discovery is that the labeled data are very limited, since they are very expensive and time-consuming to obtain. As a result, how to minimize the number of labeled data needed for effective molecular property prediction has long been a challenge in drug discovery. 

One promising direction is multi-task learning, which tries to train multiple tasks (or properties) simultaneously so that the supervision or knowledge can be shared across tasks. Indeed, multi-task learning has been successfully applied to different domains and applications such as natural language understanding~\cite{sogaard-goldberg-2016-deep,wang2021gradient}, computer vision~\cite{misra2016cross,lu2017fully}, and speech recognition~\cite{zhang2017attention,jain2018improved}. In general, the essential idea of these works is to infer the relation among tasks. For example, \cite{lu2017fully} studied the hierarchical structure of different tasks; some more recent works~\cite{master_thesis,yu2020gradient,wang2021gradient} tried to infer the pairwise relation between tasks based on the gradients or loss of the tasks. There are also some recent work on multi-task learning for molecular property prediction~\cite{dahl2014multi,ramsundar2015massively,ramsundar2017multitask,wu2018moleculenet,master_thesis,liu2019loss}, which have shown very promising results. However, drug discovery possesses certain attributes distinguishable from other domains, making it more challenging and interesting. (1) There is rich information in chemistry and biology domain, {\eg}, the task relation if we are referring molecules as data and corresponding biological effects as the tasks. Then the question is how to better utilize such domain knowledge. (2) The number of molecules for each task is comparatively small, and merging data from different tasks may lead to a severe data sparsity issue (an example in~\Cref{sec:data_generation}), which adds more obstacles for learning.

In this paper, we study multi-task learning for molecular property prediction in a different setting, where a relation graph between tasks is explicitly given via domain knowledge. We first construct a large-scale dataset called ChEMBL-STRING by combining the chemical database of bioactive molecules (ChEMBL~\cite{mendez2018chembl}) and the protein-protein interaction graph (STRING~\cite{szklarczyk2019string}). Specifically, we define a binary classification task based on an assay in ChEMBL, which measures the biological effects of molecules over a set of proteins. The relationship between different tasks are defined according to the relation of their associated sets of proteins, which can be inferred according to the protein-protein interaction graph in STRING. Finally, we are able to construct a large-scale dataset with 13,004 molecules and 382 tasks, together with the corresponding task relation graph.

With this constructed dataset, we propose a \textbf{novel} research problem: \textit{How to do structured multi-task learning with an explicit task relation graph?} Our proposed solution is SGNN-EBM, which models the structured task information in both the \textit{latent} and \textit{output} space. More specifically, a state graph neural network (SGNN) can learn effective task representations by utilizing the relation graph, where the learnt representations effectively capture the similarities between tasks in the latent space. However, given a molecule, its labels are predicted independently for each task, which ignores the task dependency, {\ie}, the dependency in the output space. Therefore, we further introduce formulating multi-task learning as structured prediction~\cite{belanger2016structured} problem, and apply an energy-based model (EBM) to model the joint distribution of the labels in the task space. Our proposed solution, coined SGNN-EBM, combines the advantages of both by adopting SGNN into the energy function in EBM, which provides higher capacity for structured task modeling. As training SGNN-EBM is generally computationally expensive, we deploy the noise contrastive estimation (NCE)~\cite{gutmann2010noise} for effective training, which trains a discriminator to distinguish the observed examples and examples sampled from a noise distribution.

Our major contributions include (1) To our best knowledge, we are the \textbf{first} to propose doing multi-task learning with an explicit task relation graph; (2) We construct a domain-specific multi-task dataset with relation graph for drug discovery; (3) We propose SGNN-EBM for task structured modeling in both the latent and output space; (4) We achieve consistently better performance using SGNN-EBM.

\section{Related Work} \label{sec:related_work}
In the multi-task learning (MTL) literature, there are two fundamental problems: (1) how to learn the relation among tasks, and (2) how to model the task relation once available. Existing works on MTL \textit{merely} focus on the first question, which can be roughly classified into two categories: architecture-specific MTL and architecture-agnostic MTL.

\textbf{Architecture-specific MTL} aims at designing special architecture to better transfer knowledge between tasks. Fully-adaptive network~\cite{lu2017fully} dynamically groups similar tasks in a hierarchical structure. Cross-stitch network~\cite{misra2016cross} applies multiple cross-stitch units and Bypass network~\cite{ramsundar2017multitask} manipulates the architecture to model task relation.
One drawback is that as the number of tasks grows, the requirement of computation memory increases linearly, which limits their application to large-scale setting (w.r.t. the task number).

\textbf{Architecture-agnostic MTL} provides a more general solution by learning to balance the tasks numerically. It has two components: a shared representation module and multiple task-specific prediction modules. Based on this framework, several methods have been proposed to learn a global linear task coefficient according to the optimization process, such as the the uncertainty~\cite{kendall2018multi}, and task gradients and losses~\cite{chen2018gradnorm,liu2019end,liu2019loss}. The learnt linear vector is then applied on the task-specific predictors. Instead of learning such linear vector, one alternative approach is to learn the pairwise task relation. RMTL~\cite{master_thesis} first handles this by applying a reinforcement learning framework to reduce the gradient conflicts between tasks.
PCGrad and GradVac~\cite{yu2020gradient,wang2021gradient} follow the same motivation and use gradient projection. However, there is one drawback on the high computational cost, since the pair-wise computation grows quadratically with the number of tasks; thus they are infeasible for large-scale setting (w.r.t. the task number).

\textbf{Molecular property prediction} has witnessed certain successful applications with MTL~\cite{kaggle2012merck,dahl2014multi,unterthiner2014deep,ramsundar2015massively,wu2018moleculenet,master_thesis,liu2018practical,liu2019loss} in terms of the robust performance gain. Furthermore, \cite{lee2019silico} finds that similarity within a target group significantly affects the performance of MTL on molecular binding prediction, revealing the importance of utilizing the task relation in drug discovery. However, all the aforementioned MTL methods do not possess the knowledge of the task relation and thus the main focus is to learn it in an architecture-specific or architecture-agnostic manner. While in this work, the task relation is given, and our focus moves to how to better model the structured task information in the MTL setting.

\section{Problem Definition \& Preliminaries} \label{sec:background}

\subsection{Problem Definition} \label{sec:task_relation_graph}
\textbf{Molecular Graph and Property Prediction.}
In molecular property prediction tasks, each data point $\vx$ is a molecule, which can be naturally viewed as a \textit{topological graph}, where atoms and bonds are nodes and edges accordingly. For each molecule $\vx$, we want to predict $T$ biological or physical \textit{properties}~\cite{wu2018moleculenet}, where each property corresponds to one \textit{task}. For notation, we want to predict $\vy=\{y_0, y_1, ..., y_{T-1} \}$ for each molecule $\vx$. Each task corresponds to $C$ classes if it is a classification problem; and specifically in this work, we will be targeting at the binary tasks, {\ie}, $C=2$ and $y_i\in \{0,1\}, \forall i\in \{0, 1, \cdots, T-1\}$. 

\textbf{Multi-Task Learning (MTL).}
Due to the inherent data scarcity issue in drug discovery~\cite{ramsundar2015massively,wu2018moleculenet,mayr2018large,hu2019pre}, training an independent model for each task often yields inferior performance. In practice~\cite{mayr2018large}, a more effective and widely-adopted approach is \textit{multi-task learning (MTL)}, which tries to optimize multiple tasks simultaneously.

\textbf{Task Relation Graph.}
A \textit{task relation graph} is $\gG=(V, E)$, where $V$ is the node set of tasks and $E$ are the corresponding edges between tasks. Here we add a linkage between two tasks if they are closely related. Thus, this relation graph can effectively complement the information sparsity of the labeled data for different tasks. More information on the task relation graph $\gG$ will be introduced in~\Cref{sec:data_generation}. 

\textbf{Structured Task Modeling.}
In this paper, we propose a \textit{novel} research problem for MTL: how to do \textit{structured task modeling} when the task relation graph is explicitly provided. Specifically, given a molecular graph $\vx$, our goal is to jointly predict its labels for $T$ tasks $\vy=\{y_0, y_1, ..., y_{T-1} \}$ with a task relation graph $\gG$. In other words, we aim to model $p(\vy|\vx, \gG)$.

\subsection{Preliminaries} \label{sec:preliminaries}
 
\textbf{Graph Neural Network (GNN)} is a powerful tool in modeling structured data, like molecular graph and task relation graph. \cite{gilmer2017neural} first proposes a general GNN framework called \textit{message passing neural network (MPNN)}. Following this, recent works have explored how to model the complex structured data like molecular graph~\cite{duvenaud2015convolutional,rong2020self,NIPS2019_9054,demirel2021analysis,ying2021transformers} and knowledge graph~\cite{kipf2016semi,xu2018powerful}. Typically for the node-level prediction, GNN models predict the node labels independently, and this limits the learning power of GNN to model the joint distribution of labels.

\textbf{Energy-Based Model (EBM)} uses a parametric energy function $E_\phi(\vx,\vy)$ to fit the data distribution~\cite{lecun2006tutorial}.
The energy function induces a density function with the Boltzmann distribution. Formally, the probability of $p_\phi(\vy|\vx)$ can be written as:
\begin{equation} \label{eq:ebm_likelihood}
\small{
\begin{aligned}
p_\phi(\vy|\vx) & = \frac{\exp(-E_\phi(\vx,\vy))}{Z_\phi(\vx)},
\end{aligned}
}
\end{equation}
where $E_\phi(\vx, \vy)$ is the energy function, with which EBM is allowed to model the structured output space. $Z_\phi(\vx) = \sum_{\vy' \in \mathcal{Y}} \exp (-E_\phi(\vx,\vy'))$ is the partition function. Here $\mathcal{Y}=\{0,1\}^{T}$ is the label space, and the partition function is computationally intractable due to the high cardinality in $ |\mathcal{Y}| = 2^{T}$. We will discuss how to cope with this issue for learning and inference in \Cref{sec:method}.

\vspace{-1ex}
\section{Dataset with Explicit Task Relation} \label{sec:data_generation}
\vspace{-1ex}
In this section, we describe ChEMBL-STRING construction, a molecular property prediction dataset together with an explicit task relation graph. The \textit{task} here refers to a binary classification problem on a ChEMBL assay~\cite{mendez2018chembl}, which measures certain biological effects of molecules, {\eg}, toxicity, inhibition or activation of proteins or whole cellular processes, etc. We focus on tasks that target at proteins ({\ie}, the binding affinity-related tasks), since the existing protein-protein interaction (PPI) data source can serve for the task relation extraction.

Our ChEMBL-STRING dataset is based on the \textit{Large Scale Comparison (LSC)} dataset proposed by~\cite{mayr2018large}, which is filtered from the ChEMBL-20 database~\cite{mendez2018chembl}. We account for a subset of 725 tasks which are protein-targeting. For each of these tasks, we collect the UniProt IDs \cite{2018uniprot} of the targeted proteins and combine all of them into a UniProt ID set. We then query the STRING database~\cite{szklarczyk2019string} to obtain PPI scores for all pairs of proteins in the set. With the collected PPI scores, we then heuristically define the edge weights~$w_{ij}$, {\ie}, task relation score, for task~$t_i$ and~$t_j$ in the task relation graph to be $\max\{\mathrm{PPI}(s_i, s_j): s_i \in S_i, s_j \in S_j\}$, where~$S_i$ denotes the protein set of task~$t_i$. Therefore, the task relation graph proposed has a high quality to reveal the actual pharmaceutical effects for the molecular drugs.

\begin{table}[htb!]
\small
\setlength{\tabcolsep}{9pt}
\centering
\caption{
% \small{
Statistics about ChEMBL-STRING datasets with explicit task relation, filtered by 3 thresholds. Threshold means the number of non-missing labels for each molecule/task.
% }
\vspace{-0.2cm}
}
\begin{tabular}{l r r r r}
\toprule
Threshold & \# Molecules & \# Tasks & Sparsity \\
\midrule
10  & 13,004 & 382 & 5.76\% \\
50  & 932    & 152 & 66.70\%\\
100 & 518    & 132 & 92.87\%\\
\bottomrule
\end{tabular}
\label{tab:chembl_statistics}
\end{table}

\begin{figure*}[t]
\centering
\includegraphics[width=0.8\textwidth]{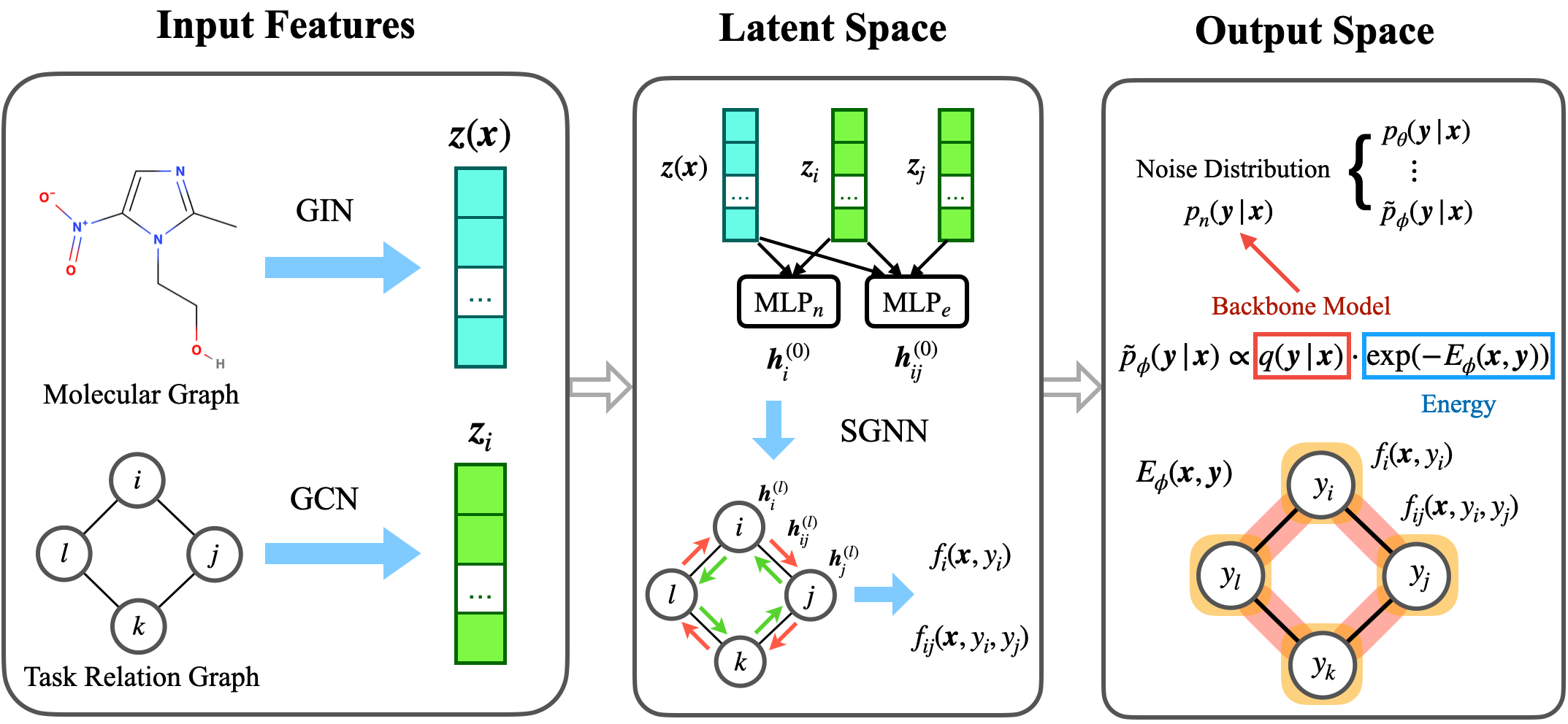}
\caption{
Pipeline of SGNN-EBM. We first obtain molecule and task embedding via GIN and GCN. Then, they are used to learn the latent representation for each task via a GNN model in the latent space. In SGNN-EBM, an SGNN model is used to model the task relation graph in the latent space and EBM learns the task distribution in the output space. The likelihood also applies the energy tilting term, which takes the same empirical distribution as the noise distribution for NCE.
}
\label{fig:pipeline}
\end{figure*}

As the experiment-based LSC dataset is very sparsely-labeled - only 0.78\% of elements of the molecule-task matrix have a label of \textit{active} or \textit{inactive}, we densify the molecule-task label matrix by iteratively filtering out molecules and tasks whose number of labels is lower than a certain threshold. By setting the threshold value to 10, 50 and 100, we obtain 3 benchmark datasets with different level of data sparsity. The statistics of the benchmark datasets are listed in \Cref{tab:chembl_statistics}, and more detailed dataset generation procedure can be found in \Cref{sec:appendix_dataset}.

\section{Method: Structured Task Modeling} \label{sec:method}
\vspace{-1ex}
%%%%%%%%%%%%%%%%%%%%%%%%%%%%%%%%%%%%%%%%%%%%%%%%%%%%%%%%%%%%
\subsection{Overview}
\vspace{-1ex}
The mainstream multi-task learning (MTL) methods~\cite{kendall2018multi,chen2018gradnorm,liu2019loss,yu2020gradient} typically learn the task relation implicitly, which can guide to balance tasks during training. While in this paper, we focus on a novel setting where the \textit{task relation graph} is explicitly given and the goal is to better model such relation graph. We first propose a dataset with an explicit task relation graph in~\Cref{sec:data_generation}, then in this section, we introduce two \textit{structured MTL} approaches to modeling the task relation in the \emph{latent} and \emph{output} space respectively.

In the latent space, we propose to learn effective task representations with a State GNN (SGNN) on the task relation graph so that the learnt representations can capture the similarity between tasks. The property $y_i$ in each task $i$ can be independently predicted with the molecule information and its own task representation. More specifically, we can define the distribution as:
\begin{equation} \label{eq:sgnn_prediction}
\small{
p_\theta(\vy|\vx, \mathcal{G})=\prod_{i=0}^{T-1} p_\theta(y_i|\vx, \mathcal{G}),
}
\end{equation}
where $p_\theta(y_i|\vx, \mathcal{G})$ is the prediction on the $i$-th task.
We present this method in~\cref{sec:method_gnn}. More detailed description of GNN can be found in~\Cref{sec:app_gnn}.

One limitation of the SGNN is that it ignores the dependency between task labels $y_i$. To handle this issue, we further propose to model the task dependency in the output space and solve it under the energy-based model (EBM) framework, as a structured prediction problem. The joint distribution of $\vy$ can be modeled with EBM as:
\begin{equation}
\small{
p_\phi(\vy|\vx, \mathcal{G}) = \frac{ \exp(-E_\phi(\vx, \vy; \mathcal{G}))}{Z_\phi},
}
\end{equation}
where $E_\phi(\vx, \vy; \mathcal{G})$ is the energy function with flexible format. The noise contrastive estimation (NCE) is used to learn the EBM efficiently, and an outline of these methods is depicted in~\Cref{fig:pipeline}.

Then we combine the advantages of both approaches by accounting the SGNN for energy function in EBM. Thus we are able to model the task relation in both the latent and output space, and we name this method as \textbf{SGNN-EBM} for solving structured MTL problems.

%%%%%%%%%%%%%%%%%%%%%%%%%%%%%%%%%%%%%%%%%%%%%%%%%%%%%%%%%%%%
\vspace{-1ex}
\subsection{Modeling Task Relation in Latent Space} \label{sec:method_gnn}
\vspace{-1ex}
We propose \textit{State GNN (SGNN)} to model the task relation in the latent space. The task relation is implicitly encoded in the learnt representations, and the final predictions are made independently for each task. We illustrate the pipeline of this model as follows.

\textbf{Node- and Edge-Level Inputs.} We first encode the molecules and tasks into the \textit{embedding} space. For molecules, we adopt graph isomorphism network (GIN)~\cite{xu2018powerful}, and the molecule embedding is $\vz(\vx) \in \mathbb{R}^{d_m}$, where $d_m$ is the embedding dimension. Then for tasks, we use one-hot encodings (w.r.t. the task index) and pass them through a graph convolutional network (GCN)~\cite{kipf2016semi} to get task embedding: $\vz^{(i)} \in \mathbb{R}^{d_e}, \forall i \in \{0, 1, \hdots, T-1\}$, where $d_t$ is the task embedding dimension. More details of GIN and GCN can be found in~\Cref{sec:app_gin,sec:app_gcn}. Given the molecule and task embeddings, we will use them to construct the node- and edge-level inputs to SGNN as:
\begin{equation} \label{eq:gnn_node_edge_inputs}
\small{
\begin{aligned}
& \vh_i^{(0)}(\vx) = \text{MLP}^{(0)}_n(\vz(\vx) \oplus \vz^{(i)})\\
& \vh_{ij}^{(0)}(\vx) = \text{MLP}^{(0)}_e(\vz(\vx) \oplus \vz^{(i)} \oplus \vz^{(j)}),
\end{aligned}
}
\end{equation}
where $\oplus$ is the concatenation of two tensors. $\text{MLP}^{(0)}_n: \sR^{d_m+d_t}\rightarrow \sR^{C\times d}$ and $\text{MLP}^{(0)}_e: \sR^{d_m+2d_t}\rightarrow \sR^{C\times C\times d}$ are two multi-layer perceptron (MLP) layers, operating on the node- and edge-level respectively. $d$ is the dimension of the latent representation and $C=2$ is the class number, and it also represents the states on each node and edge in SGNN. The node- and edge-level inputs in~\Cref{eq:gnn_node_edge_inputs} will then be fed to SGNN.

\textbf{State GNN (SGNN).} Different from the mainstream GNN models, SGNN has $C$ and $C\times C$ states on each node and edge respectively, where each state delegates the representation for the corresponding label. Concretely, every node state represents the task w.r.t. the corresponding label, and edge state is composed of the pair-wise states from the two endpoint nodes. Thus, the representation for each node and edge state is defined as:
\begin{equation}
\small{
\begin{aligned}
& \vh_i^{(0)}(\vx, y_i) = \vh_i^{(0)}(\vx)[y_i]\\
& \vh_{ij}^{(0)}(\vx, y_i, y_j) = \vh_{ij}^{(0)}(\vx)[y_i,y_j].
\end{aligned}
}
\end{equation}
In this way, the representations of nodes and edges can well capture the information of each node label as well as the pairwise labels on an edge.

Such state-level view builds up the smallest granularity in SGNN. For example, during \textbf{message-passing propagation}, the key function in SGNN, only information with the same state will be exchanged between nodes and edges. Specifically, the propagation on the $l$-th layer is:
\begin{equation}\label{eq:gnn_message_passing}
\small{
\begin{aligned}
& \vh_i^{(l+1)}(\vx, y_i) = \\
& \quad \quad \text{MPNN}_n^{(l+1)}\Big(\vh_i^{(l)}(\vx, y_i), \big\{ \vh_{ij}^{(l)}(\vx, y_i, y_j) \mid \forall j, y_j \big\} \Big)\\
& \vh_{ij}^{(l+1)}(\vx, y_i, y_j) = \\
& \quad \quad \text{MPNN}_e^{(l+1)}\Big(\vh_i^{(l)}(\vx, y_i), \vh_j^{(l)}(\vx, y_j), \vh_{ij}^{(l)}(\vx, y_i, y_j)\Big),
\end{aligned}
}
\end{equation}
where MPNN stands for the message-passing neural network layer~\cite{gilmer2017neural}. $\text{MPNN}_n$ is doing node aggregation by gathering information from edges with the same node state $y_i$; $\text{MPNN}_e$ stores the messages for each state pair ($y_i, y_j$) with the corresponding state information from the nodes. After repeating~\Cref{eq:gnn_message_passing} $L$ times, we obtain the latent representation for each task given the molecule.

\textbf{Independent Label Prediction.}
Finally, we make predictions for each task independently as~\Cref{eq:sgnn_prediction}. For each task $i$, we first get the node representation by concatenating the two state representations, after which we apply a readout function $R$:
\begin{equation} \label{eq:gnn_readout_function}
\small{
\begin{aligned}
f_i(\vx) & = R(\{\vh^{(l)}_i(\vx, 0) \oplus \vh^{(l)}_i(\vx, 1) \mid l=1, \hdots, L\}),
\end{aligned}
}
\end{equation}
where $R: \sR^{2dL}\rightarrow \sR$ is an MLP layer.
Because $C=2$ is the binary classification, the label distribution is defined via a sigmoid function, {\ie}, $p(y_i=1 | \vx, \mathcal{G}) = \text{sigmoid}(f_i(\vx))$. The loss function is the binary cross entropy function over all $T$ tasks:
\begin{equation}
\small{
\gL = \sum_{i=0}^{T-1} \log p(y_i | \vx, \mathcal{G}).
}
\end{equation}
Despite the effectiveness of learning task representations, SGNN fails to directly model the task dependency when making predictions as different task labels are predicted separately. To address this issue, we next propose a general method for modeling the task dependency under the structured prediction framework, which is able to predict task labels collectively to improve the result.

%%%%%%%%%%%%%%%%%%%%%%%%%%%%%%%%%%%%%%%%%%%%%%%%%%%%%%%%%%%%
\vspace{-1ex}
\subsection{Modeling Task Relation in Output Space} \label{sec:method_ebm}
\vspace{-1ex}
The aforementioned MTL methods are predicting each task independently. However, there also exists a task distribution in the output space, {\ie}, $p(\vy=y_0, y_1, \hdots, y_{T-1}|\vx)$. In this subsection, we propose to apply an energy-based model (EBM) to inject the prior knowledge about task dependency and model it with joint task distribution.

We define the \textbf{energy function} as the summation of first-order (node) and second-order (edge) factors on the graph:
\begin{equation} \label{eq:energy_function}
\small{
\begin{aligned}
E_\phi(\vx,\vy) = - \sum_{i=0}^{T-1} f_i(\vx,y_i) - \lambda \sum_{\langle i, j \rangle \in \mathcal{G}}  f_{ij}(\vx,y_i,y_j),
\end{aligned}
}
\end{equation}
where $\lambda$ is a weighting coefficient. Thus the conditional probability under the EBM framework is defined as:
\begin{equation} \label{eq:energy_probability}
\small{
p_\phi(\vy|\vx) = \frac{\exp\Big( \sum_i f_i(\vx,y_i) + \sum_{ij} f_{ij}(\vx,y_i,y_j) \Big)}{Z_\phi}.
}
\end{equation}

\textbf{Activation Function.}
We apply the activation function $\sigma(\cdot) = \log (\text{softmax}(\cdot))$ on the first- and second-order factors. Then the readout function is $\tilde{R}(\cdot) =  \log (\text{softmax}(\text{MLP}( \cdot )))$, where the softmax function is applied on the label/state space of each task and each task pair. The softmax function normalizes the scores of different label candidates, allowing us to compare them in the same range between 0 and 1. The logarithm function further scales the energy to 0 to $\infty$, which is a common practice in EBM.

\textbf{Energy Tilting Term.}
We have introduced EBM to model task relations in output space. However, directly training the energy-based model is still a challenging problem. To alleviate this issue, we leverage the energy tilting term from~\cite{dai2014generative,xie2016theory,nijkamp2020learning,arbel2020generalized}, which takes EBM in the form of a correction or an exponential tilting of a pre-trained backbone model $q(\vy|\vx)$. The pre-trained backbone model acts as a base model, and the energy function $\exp(-E_{\phi}(\vx,\vy))$ tries to tilt the base model for better results, yielding an integrated model as: $\tilde{p}_{\phi}(\vy|\vx) \propto q(\vy|\vx) \cdot \exp(-E_{\phi}(\vx,\vy))$,
where the integrated model $\tilde{p}_{\phi}(\vy|\vx)$ is named the \textit{energy tilting distribution}. We will illustrate how to combine this energy tilting term in the learning and inference below.

%%%%%%%%%%%%%%%%%%%%%%%%%%%%%%%%%%%%%%%%%%%%%%%%%%%%%%%%%%%%
\vspace{-1ex}
\subsection{SGNN-EBM} \label{sec:sgnn_ebm}
\vspace{-1ex}
Then we will combine the structured modeling on both latent and output space together. As mentioned before, the energy function in EBM can have flexible formulation~\cite{lecun2006tutorial}; thus, we may as well parameterize it by adopting the node- and edge-level representation from SGNN. With minor modifications we have:
\begin{equation} \label{eq:gnn_energy_function}
\small{
\begin{aligned}
& f_i(\vx,y_i) = \tilde{R}(\{\vh^{(l)}_i(\vx, y_i) \mid l=1, \hdots, L\})\\
& f_{ij}(\vx,y_i,y_j) = \tilde{R}(\{\vh^{(l)}_{ij}(\vx, y_i, y_j) \mid l=1, \hdots, L\}),
\end{aligned}
}
\end{equation}
where $\tilde{R}:\sR^{dL}\rightarrow \sR$ is a readout function defined as $\tilde R = \sigma(\text{MLP}( \cdot ))$ and $\sigma(\cdot)$ is the activation function. \Cref{eq:gnn_energy_function} is mapping the node and edge representations to scalars (or energies) indiced with the corresponding node and edge label.

As the number of message-passing layers $L$ increases, the SGNN-based energy function (\Cref{eq:gnn_energy_function}) can be seen as a general form to capture the higher-order dependency. However, according to the energy function decomposition in~\Cref{eq:energy_function}, only first- and second-order factors are considered during the EBM learning and inference. This discrepancy may raise some potential concern, and we carry on an ablation study in~\Cref{sec:effect_of_gnn_layer}, where we empirically prove that slightly increasing $L$ can be beneficial for the generalization performance. Yet, this is still worth further exploration in the future.

In the following sections, we will introduce how to do NCE learning and Gibbs sampling inference for our proposed SGNN-EBM model.

%%%%%%%%%%%%%%%%%%%%%%%%%%%%%%%%%%%%%%%%%%%%%%%%%%%%%%%%%%%%
\vspace{-1.5ex}
\subsubsection{Learning} \label{sec:ebm_training_nce}
\vspace{-1.5ex}
The learning process aims at optimizing $\phi$ to maximize the data likelihood. However, the problem is nontrivial as the partition function $Z_\phi$ is intractable. Our approach addresses this by using noise contrastive estimation (NCE)~\cite{gutmann2010noise}, which casts the problem of maximizing log-likelihood into a contrastive learning task. We first take the normalization constant $Z_\phi$ in \Cref{eq:ebm_likelihood} as a learned scalar parameter. Then we transform the EBM learning into a binary classification problem by maximizing the following objective:
\begin{equation}\label{eq:nce}
\small{
\begin{aligned}
\gL_{NCE} & =
\mathbb{E}_{\vy \sim p_n} \log \frac{p_n(\vy|\vx)}{p_n(\vy|\vx)+ p_{\phi}(\vy|\vx)}\\
& + \mathbb{E}_{\vy \sim p_{\text{data}}} \log \frac{p_\phi(\vy|\vx)}{p_n(\vy|\vx) + p_{\phi}(\vy|\vx)},
\end{aligned}
}
\end{equation}
where $p_{\text{data}}$ is the underlying data distribution, $p_{\phi}$ is the model distribution to approximate data distribution, and $p_n$ is a noise distribution, whose samples serve as negative examples in the contrastive learning objective. Ideally, $p_{\phi}$ will be trained to approximate $p_{\text{data}}$ for any noisy distribution. Yet in practice, the noise distribution should be close to the data distribution to facilitate the mining of hard negative samples. In addition~\cite{mnih2012fast}, given an expressive energy function, we can fix $Z_\phi=1$ and the resulting learned EBM will be self-normalized.

\begin{table*}[t]
\small
\setlength{\tabcolsep}{20pt}
\centering
\caption{
% \small{
Main MTL results. All datasets are split into 8-1-1 for train, valid, and test respectively. For each method, we run 5 seeds and report the mean and standard deviation. The best performance is \textbf{\underline{highlighted}}.
% }
\vspace{-0.2cm}
}
\begin{adjustbox}{max width=\textwidth}
\begin{tabular}{l l c c c}
\toprule
Method & $p_n$ & ChEMBL 10& ChEMBL 50& ChEMBL 100\\
\midrule
STL & -- & 71.67 $\pm$ 0.64 & 73.57 $\pm$ 1.20& 70.81 $\pm$ 1.28\\
MTL & -- & 74.83 $\pm$ 0.61 & 79.37 $\pm$ 1.76& 77.78 $\pm$ 1.59\\
UW & -- & 72.49 $\pm$ 0.53 & 79.68 $\pm$ 0.98 & 78.71 $\pm$ 1.93\\
GradNorm & -- & 75.17 $\pm$ 0.77 & 79.46 $\pm$ 1.27& 78.75 $\pm$ 1.60\\
DWA & -- & 72.45 $\pm$ 1.31 & 79.35 $\pm$ 0.68& 78.21 $\pm$ 2.31\\
LBTW & -- & 75.21 $\pm$ 0.49 & 79.52 $\pm$ 0.56 & 79.07 $\pm$ 0.99\\
\midrule
SGNN &-- & 77.90 $\pm$ 0.88& 79.67 $\pm$ 0.87& 80.19 $\pm$ 0.67\\
SGNN-EBM & SGNN (Fixed) &  78.04 $\pm$ 0.73 & 80.34 $\pm$ 1.08 & 80.48 $\pm$ 1.93\\
SGNN-EBM & SGNN (Adaptive) & \textbf{\underline{78.35 $\pm$ 1.07}} & \textbf{\underline{80.54 $\pm$ 1.02}} & \textbf{\underline{81.15 $\pm$ 0.59}}\\

\bottomrule
\end{tabular}
\end{adjustbox}
\label{tab:complete_results}
\end{table*}

\textbf{NCE with Tilting Term.}
The above objective function seems complicated. Nevertheless, it will become more concise as we combine the energy tilting term into NCE learning. We apply the backbone model for the noise distribution, {\ie}, $p_n=q$, and replace the energy tilting term into~\Cref{eq:nce}. With the self-normalized partition function, the NCE learning with energy tilting term can be written as:
\begin{equation}\label{eq:nce_tilting}
\small{
\begin{aligned}
\tilde \gL_{NCE} & = \mathbb{E}_{\vy \sim p_n} \log \frac{1}{1 + \exp(-E_\phi(\vx, \vy))} \\
& + \mathbb{E}_{\vy \sim p_{\text{data}}} \log \frac{1}{1 + \exp(E_\phi(\vx, \vy))}.
\end{aligned}
}
\end{equation}
In this new objective function, we only need to draw samples from the noise distribution without computing their density, which is easy to operate. More detailed derivations are attached in~\Cref{sec:appendix_energy_tilting_term}.

\textbf{The Choice of Noise Distribution.}
One key component in NCE training is the choice of the noise distribution, $p_n$. NCE works for any given noise distribution, yet the algorithm empirically converges faster if the noise distribution $p_n$ can stay close to the model distribution $p_\phi$~\cite{song2021train}. In the experiment, we propose two options for selecting the noise distributions. (1) We use a pre-trained model to be a \textit{fixed} noise distribution, {\eg}, the SGNN proposed in~\Cref{sec:method_gnn} and $p_n = p_{\theta}$. (2) We adopt an \textit{adaptive} noise distribution, and start with a pre-trained model. The difference is that after training with this pre-trained noise distribution for a few epochs, we will gradually update the noise distribution with our learned model, {\ie}, updating $p_n$ with the latest $\tilde{p}_\phi$. The second idea aligns well with the curriculum learning~\cite{bengio2009curriculum}, a learning process starting with easy data to hard data. Thus another way to interpret the adaptive noise distribution is that, we start with a simple distribution (from a pre-trained model distribution) and gradually using harder distribution (from the latest model distribution). We investigate the effect on the choices of noise distributions for NCE learning in the ablation study in~\Cref{sec:effect_of_noise_distribution}.

\textbf{Imputation for Missing Labels.}
For the SGNN-EBM training proposed in~\Cref{sec:sgnn_ebm}, we use the task distribution for predicting each data point, $p_\phi(\vy|\vx)$, but some tasks do not have valid labels due to the label sparsity, as discussed in~\Cref{sec:intro,sec:data_generation}. In SGNN-EBM, we propose to use the backbone model, $q$, to fill in the missing labels so as to calculate the probability. This strategy shares similar idea to the EM algorithm~\cite{neal1998view}, which allows us to maximize a variational lower bound of the data likelihood. Empirically, experiment results help support this imputation strategy, yet, this is still work investigating in the future.

%%%%%%%%%%%%%%%%%%%%%%%%%%%%%%%%%%%%%%%%%%%%%%%%%%%%%%%%%%%%
\vspace{-1.5ex}
\subsubsection{Inference} \label{sec:ebm_inference}
\vspace{-1.5ex}
The inference procedure aims at computing the marginal distribution for each task, which can be further utilized for the label prediction for each task. The main challenge is how to calculate the intractable partition function during inference. We propose to approximate the distribution via Gibbs sampling~\cite{geman1984stochastic}. Gibbs sampling is a classic MCMC-based inference method and the core idea is to generate samples by sweeping through each variable to a sample with the remaining variables fixed.

To adopt Gibbs sampling in our setting, for each data and $T$ labels, $(\vx, y_0, \hdots, y_{T-1})$, we iteratively sample label for each task with other labels fixed. The update function at each iteration is:
\begin{equation}
\small{
\begin{aligned}
& p_\phi(y_i| \vy_{-i}, \vx) \\
= & \frac {\exp\big( f_i(\vx,y_i) + \sum_{\langle i,j \rangle \in \mathcal{G}} f_{ij}(\vx,y_i,y_j) \big)}
{\sum_{y_i=0}^{C-1} \exp \big( f_i(\vx,y_i) + \sum_{\langle i,j \rangle \in \mathcal{G}} f_{ij}(\vx,y_i,y_j) \big) }, 
\end{aligned}
}
\end{equation}
where $\vy_{-i}$ denotes all $T$ task labels except the $i$-the task. Then we take this as the tilting term, and apply $\tilde{p}(\vy|\vx) = p_\phi(\vy|\vx) \cdot q(\vy|\vx)$ for sampling. To accelerate the convergence of Gibbs sampling, we take the backbone model for initial distribution.

\begin{table*}[htb!]
\small
\setlength{\tabcolsep}{18pt}
\centering
\caption{
\small{
The effect of different noise distributions $p_n$ in NCE. Here all the noise distributions are fixed.
}
\label{tab:effect_of_noise_distribution}
\vspace{-0.3cm}
}
\begin{adjustbox}{max width=\textwidth}
\begin{tabular}{l l c c c}
\toprule
Method & $p_n$ & ChEMBL-STRING 10 & ChEMBL-STRING 50 & ChEMBL-STRING 100\\
\midrule
MTL& -- & 74.83 $\pm$ 0.61 & 79.37 $\pm$ 1.76& 77.78 $\pm$ 1.59\\
UW& -- & 72.49 $\pm$ 0.53 & 79.68 $\pm$ 0.98 & 78.71 $\pm$ 1.93\\
GradNorm& -- & 75.17 $\pm$ 0.77 & 79.46 $\pm$ 1.27& 78.75 $\pm$ 1.60\\
DWA & -- & 72.45 $\pm$ 1.31 & 79.35 $\pm$ 0.68& 78.21 $\pm$ 2.31\\
LBTW & -- & 75.21 $\pm$ 0.49 & 79.52 $\pm$ 0.56 & 79.07 $\pm$ 0.99\\
SGNN &-- & 77.90 $\pm$ 0.88& 79.67 $\pm$ 0.87& 80.19 $\pm$ 0.67\\

\midrule
SGNN-EBM & Uniform & 58.66 $\pm$ 4.65 & 73.55 $\pm$ 0.61& 75.49 $\pm$ 1.64\\
SGNN-EBM & MTL & 75.71 $\pm$ 0.41 & 79.96 $\pm$ 1.41 & 78.41 $\pm$ 1.37\\
SGNN-EBM & UW & 74.36 $\pm$ 0.87 & 80.26 $\pm$ 0.67& 79.12 $\pm$ 1.79\\
SGNN-EBM & GradNorm & 75.83 $\pm$ 0.73 & 80.18 $\pm$ 1.04 & 79.34 $\pm$ 1.31\\
SGNN-EBM & DWA & 75.22 $\pm$ 1.16 & 80.18 $\pm$ 0.74 & 79.01 $\pm$ 1.94\\
SGNN-EBM & LBTW & 76.16 $\pm$ 0.54 & 80.04 $\pm$ 0.50 & 79.68 $\pm$ 0.93\\
SGNN-EBM & SGNN & 78.04 $\pm$ 0.73 & 80.34 $\pm$ 1.08 & 80.48 $\pm$ 1.93\\
\bottomrule
\end{tabular}
\end{adjustbox}
\end{table*}

\vspace{-1ex}
\section{Experiment Results} \label{sec:experiment}
\vspace{-1ex}
%%%%%%%%%%%%%%%%%%%%%%%%%%%%%%%%%%%%%%%%%%%%%%%%%%%%%%%%%%%%
\subsection{Main Results} \label{sec:empirical_results}
\vspace{-1ex}
\textbf{Baselines.}
As described in~\Cref{sec:related_work}, the memory cost of architecture-specific MTL methods ({\eg}, Bypass network) is $O(T)$, and pair-wise architecture-agnostic MTL methods (RMTL~\cite{master_thesis}, PCGrad~\cite{yu2020gradient}, GradVac~\cite{wang2021gradient}) have $O(T^2)$ time complexity. Both are infeasible in the large-scale MTL setting (w.r.t. the number of tasks), so we exclude them in the experiments. For the baseline methods, we include standard single-task learning (STL), standard multi-task learning (MTL), Uncertainty Weighing (UW)~\cite{kendall2018multi}, GradNorm~\cite{chen2018gradnorm}, Dynamic Weight Average (DWA)~\cite{liu2019end}, and Loss-Balanced Task Weighting (LBTW)~\cite{liu2019loss}.

\textbf{Our Methods.}
We first test SGNN, which \textit{only} models the task relation graph in the latent space.
On the other hand, EBM is very sensitive to the noise distribution, leading to unstable performance. Thus we will not test it separately as SGNN, and two following ablation studies can reveal more insights for it. Then we test our main proposal, SGNN-EBM. SGNN-EBM models the task relation graph in both the latent and output space under the EBM framework, where the energy function is defined as the SGNN. We explore two noise distributions in the NCE learning steps: (2.1) the first is a fixed pre-trained SGNN, $p_n=p_\theta$; (2.2) the second is taking the pre-trained SGNN, $p_n=p_\theta$, as initial noise distribution, and then adaptively updating this noise distribution with the latest model distribution $p_n=\tilde{p}_\phi$. More training details can be found in~\Cref{sec:appendix_training_details}.

\textbf{Evaluation.}
We follow the mainstream evaluation metrics on MTL for drug discovery, {\ie}, the mean of ROC-AUC over all $T$ tasks. ROC-AUC is ranking-based, thus it can better match with the class-imbalance settings like molecular property prediction in drug discovery.

\textbf{Observation.}
We adopt the proposed dataset with three thresholds introduced in~\Cref{sec:data_generation} for experiments. The main results are in~\Cref{tab:complete_results}. First we can see all the MTL methods are better than the STL, which matches with the common acknowledgement that the joint learning can improve the overall performance. Then for our proposed methods, we can see that modeling task relation in the latent space using SGNN reaches a good performance compared to all MTL baselines, while combining it with the EBM in the output space, {\ie}, SGNN-EBM, can reach the best performance on all datasets. For the two SGNN-EBM models, they are consistently better than the SGNN model, while adaptively updated noise distribution can reach best performance. All these observations deliver an important message: structured task modeling is useful in MTL, and SGNN-EBM is an effective solution in achieving this goal.

%%%%%%%%%%%%%%%%%%%%%%%%%%%%%%%%%%%%%%%%%%%%%%%%%%%%%%%%%%%%
\vspace{-1ex}
\subsection{Ablation Study 1: The Effect of $p_n$} \label{sec:effect_of_noise_distribution}
\vspace{-1ex}
In the NCE learning of EBMs, the performance highly depends on the noise distribution $p_n$. In~\Cref{tab:complete_results} we show that the best method is SGNN-EBM with SGNN as both the energy function and noise distribution. Indeed we can take one uniform distribution and all pre-trained models (prior distribution) as the noise distribution, and we show that NCE-based structured prediction can obtain consistent performance gain when comparing to the corresponding prior distribution.

As in~\Cref{tab:effect_of_noise_distribution}, the improvement by structured prediction is not huge but consistent on all datasets: for each pre-trained model, its SGNN-EBM counterpart can consistently improve the performance by taking it as a prior distribution in NCE learning. Such consistency consolidates the effectiveness of our solution.

%%%%%%%%%%%%%%%%%%%%%%%%%%%%%%%%%%%%%%%%%%%%%%%%%%%%%%%%%%%%
\vspace{-1ex}
\subsection{Ablation Study 2: The Effect of $L$} \label{sec:effect_of_gnn_layer}
\vspace{-1ex}

\begin{table}[htb!]
\small
\setlength{\tabcolsep}{5pt}
\centering
\caption{
\small
The effect of layer number in SGNN, with 3 thresholds on ChEMBL-STRING. 
\vspace{-0.24cm}
}
\begin{adjustbox}{max width=\textwidth}
\begin{tabular}{c c c c}
\toprule
\# layer & 10 & 50 & 100\\
\midrule
0 & 77.45 $\pm$ 1.03 & 80.63 $\pm$ 0.80 & 80.82 $\pm$ 2.09\\
2 & 77.56 $\pm$ 1.00 & 80.78 $\pm$ 0.85 & 81.13 $\pm$ 2.04\\
4 & 76.98 $\pm$ 0.91 &  80.42 $\pm$ 0.82 & 81.06 $\pm$ 2.09\\
\bottomrule
\end{tabular}
\label{tab:effect_layer_number}
\end{adjustbox}
\end{table}

We test SGNN-EBM* with $L=0,2,4$ with all the other hyper-parameters fixed, where $L$ is the number of layers in GNN. In the NCE learning, we are adapting the noise distributions from a pre-trained SGNN model, $p_\theta$. The parameter $L$ reflects that each node (molecule-task) in the graph aggregates features from its $L$-hop neighborhood.

As observed in~\Cref{tab:effect_layer_number}, the SGNN-EBM improves the performance slightly with larger $L$ in SGNN owing to the ability to model longer-term dependencies among labels. However, as $L$ increases, the performance will drop instead. One possible explanation is that the inference method, Gibbs Sampling, defined in~\Cref{sec:method_ebm} only considers first- and second-order factors, thus it fails to capture the long-term dependencies.

\vspace{-1.5ex}
\section{Conclusion and Future Direction}
\vspace{-2ex}
In this paper, we propose a novel research problem of MTL for molecular property prediction with an explicit task relation graph. We propose a novel approach to modeling the task relations in both the \emph{latent} and \emph{output} space. Experimental results demonstrate that SGNN-EBM outperforms competitive baselines.

We want to highlight that SGNN-EBM can fit to broad MTL problems, as long as the explicit task relation is accessible. But as the first step along this direction, we would like to start from a modest setting with assurance from the oracle, like explicit task relation from drug discovery domain. In addition, structured task modeling opens a new and promising research venue. For example, some MTL methods (RMTL~\cite{master_thesis}, GradVac~\cite{wang2021gradient}) are able to extract the pairwise similarity to compose a task relation graph; yet, this view point is unexplored and would be interesting to combine with SGNN-EBM as the next step.

\section*{Acknowledgements}
This project is supported by the Natural Sciences and Engineering Research Council (NSERC) Discovery Grant, the Canada CIFAR AI Chair Program, collaboration grants between Microsoft Research and Mila, Samsung Electronics Co., Ltd., Amazon Faculty Research Award, Tencent AI Lab Rhino-Bird Gift Fund and a NRC Collaborative R\&D Project (AI4D-CORE-06). This project was also partially funded by IVADO Fundamental Research Project grant PRF-2019-3583139727.

\bibliography{reference}
\bibliographystyle{apalike}

\clearpage
\appendix

\thispagestyle{empty}

\onecolumn \makesupplementtitle
%%%%%%%%%% Dataset Generation %%%%%%%%%%
\section{ChEMBL-STRING Dataset Generation} \label{sec:appendix_dataset} 

We propose ChEMBL-STRING, a multi-task learning dataset with explicit task relation for the molecular property prediction. This new dataset is built on the Large Scale Comparison (LSC) dataset~\cite{mayr2018large}, and we list the three main steps in~\Cref{sec:app_dataset_step_1,sec:app_dataset_step_2,sec:app_dataset_step_3}.

\subsection{Filtering molecules} \label{sec:app_dataset_step_1}
Among 456,331 molecules in the LSC dataset, 969 are filtered out following the pipeline in~\cite{hu2019strategies}. Here we describe the detailed filtering process, and the molecules filtered out in each step.
\begin{enumerate}[noitemsep,topsep=0pt]
    \item Discard the \texttt{None}s in the compound list.
    \item Filter out the molecules with~$\leq$ 2 non-H atoms.
    \item Retain only the largest molecule in the SMILES string. \textit{E.g.} if the compound is a organic hydrochloride, say~$\mathrm{CH_3NH_3^+Cl^-}$, we retain only the organic compound after removing $\mathrm{HCl}$, in this case~$\mathrm{CH_3NH_2}$.
    \item Filter out molecules with molecular weight~$<$ 50 and 9 with molecular weight~$>$ 900.
\end{enumerate}

\subsection{Querying the PPI scores} \label{sec:app_dataset_step_2}
Then we obtain the PPI scores by quering the ChEMBL~\cite{mendez2018chembl} and STRING~\cite{szklarczyk2019string} databases. The details are as follows:
\begin{enumerate}[noitemsep,topsep=0pt]
    \item The LSC dataset~\cite{mayr2018large} gives the ChEMBL ID for each assay. We use the assay id to query the ChEMBL database by visiting \texttt{https://www.ebi.ac.uk/chembl/api/data/assay/[assay\_id]} for target ID. We then query the ChEMBL database by visiting \texttt{https://www.ebi.ac.uk/chembl/api/data/ target/[target\_id]} for UniProt~\cite{2018uniprot} information. We save all the UniProts related to each target in a list. We discard assays with no associated UniProt, and confirm that all remaining assays are targeting human proteins.
    \item Next, we query the STRING database for the corresponding STRING ID. For each UniProt, we visit \texttt{https://string-db.org/api/xml/get\_string\_ids?identifiers=[uniprot]}. We discard UniProts with no available StringIDs. The String ID list is then sent to \texttt{https://string-db.org/ api/tsv-no-header/network} via a POST request to obtain the human PPI scores.
\end{enumerate}

\subsection{Constructing the Task Relation Graph} \label{sec:app_dataset_step_3}
Finally, we calculate the edge weights~$w_{ij}$, {\ie}, task relation score, for task~$t_i$ and~$t_j$ in the task relation graph to be $\max\{\mathrm{PPI}(s_i, s_j): s_i \in S_i, s_j \in S_j\}$, where~$S_i$ denotes the protein set of task~$t_i$. The resulting task relation graph has 1,310 nodes and 9,172 edges with non-zero weights. Note that 96\% of the protein-targeted tasks only target a single protein, for which the relation score of these tasks is exactly the PPI score between their target proteins. We then densify the dataset via the following filtering process:
\begin{enumerate}[noitemsep,topsep=0pt]
    \item We filter out all isolated tasks.
    \item We define a threshold $\tau$ and iteratively filter out molecules with number of labels below  $\tau$, tasks with number of labels below $\tau$, and tasks with number of positive or negative labels below 10. We repeat this until no molecule or task is filtered out.
\end{enumerate}
The statistics of the resulting ChEMBL-STRING dataset with three thresholds can be found at~\Cref{tab:chembl_statistics}.

%%%%%%%%%% GIN %%%%%%%%%%
\section{GIN for Molecule Embedding} \label{sec:app_gin}
The Graph Isomorphism Network (GIN) is proposed in~\cite{xu2018powerful}. It was originally proposed for the simple graph structured data, where each node has one discrete label and no extra edge information is provided. Here we adopt a customized GIN from a recent paper~\cite{hu2019strategies}. With this customized GIN as the base model, plus pre-training techniques, \cite{hu2019strategies} can reach the state-of-the-art performance on several molecular property prediction tasks. Thus we adopt this customized GIN model in our work.

Following the notation in~\Cref{sec:background}, each molecule is represented as a molecular graph, {\ie}, $\vx = (X, E)$, where $X$ and $E$ are feature matrices for atoms and bonds respectively. Suppose for one molecule, we have $n$ atoms and $m$ edges. The message passing function is defined as:
\begin{equation}
z_i^{(k+1)} = \text{MLP}_{\text{atom}}^{(k+1)} \Big(z_i^{(k)} + \sum_{j \in \mathcal{N}(i)} \big( z_j^{(k)} + \text{MLP}_{\text{bond}}^{(k+1)}(E_{ij}) \big) \Big),
\end{equation}
where $z_0=X$ and $\text{MLP}_{\text{atom}}^{(k+1)}$ and $\text{MLP}_{\text{bond}}^{(k+1)}$ are the $(l+1)$-th MLP layers on the atom- and bond-level respectively. Repeating this for $K$ times, and we can encode $K$-hop neighborhood information for each atom in the molecular data, and we take the last layer for each node/atom representation. The graph representation is the mean of the node representation, {\ie}, the molecule representation in this paper:
\begin{equation}
z(\vx) = \frac{1}{N} \sum_i z_i^{(K)}
\end{equation}

%%%%%%%%%% GCN %%%%%%%%%%
\section{GCN for Task Embedding} \label{sec:app_gcn}
We use graph convolutional network (GCN)~\cite{kipf2016semi} for the task embedding. For the $i$-th task, we first get its one-hot encoding and then pass it through an embedding layer, with the output denoted as $\ve_i \in \mathbb{R}^{d_t \times 1}, \forall i \in \{0, 1, \hdots, T-1\}$, where $d_t$ is the task embedding dimension. $\mE = \{ \ve_0, \ve_1, \hdots, \ve_{T-1}\}^T \in \mathbb{R}^{T \times d_t}$ is the initial embedding matrix for $T$ tasks. Then we pass $\mE$ through a GCN and the output embedding for the $i$-th task is $\vz^{(i)} = \text{GCN}(\mE)_i, \forall i \in \{0, 1, \hdots, T-1\}$.

%%%%%%%%%% GNN %%%%%%%%%%
\section{SGNN for Modeling Latent Space} \label{sec:app_gnn}

\begin{figure}[thbp!]
\centering
\includegraphics[width=0.8\textwidth]{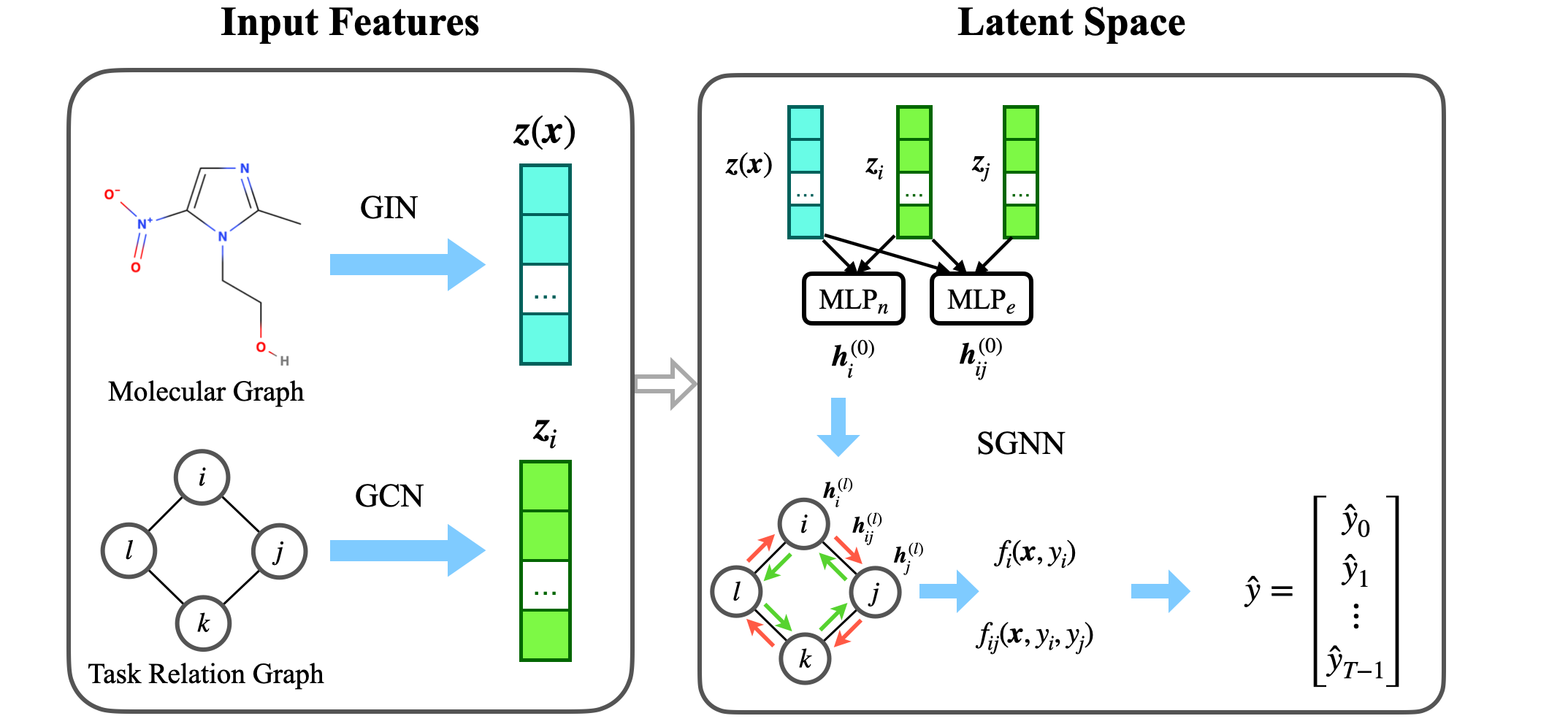}
\caption{
\small
Pipeline of GNN. We first obtain molecule and task embedding via GIN and GCN. Then they are concatenated and passed through a GNN to better learn the task representation. The final prediction for each task is predicted independently on each node representation.
}
\label{fig:pipeline_gnn}
\end{figure}

In this section, we give a detailed illustration of our proposed State GNN (SGNN) model in~\Cref{sec:method_gnn}. The general pipeline is shown in~\Cref{fig:pipeline_gnn}.

First let us quickly review the node- and edge-level inputs:
\begin{equation}
\begin{aligned}
& \vh_i^{(0)}(\vx) = \text{MLP}^{(0)}_n(\vz(\vx) \oplus \vz^{(i)})\\
& \vh_{ij}^{(0)}(\vx) = \text{MLP}^{(0)}_e(\vz(\vx) \oplus \vz^{(i)} \oplus \vz^{(j)}),
\end{aligned}
\end{equation}
and as discussed in~\Cref{sec:method_gnn}, the biggest difference between SGNN and the mainstream GNN models is that in SGNN, each node and each edge has two and four state respectively, where each state of a node/edge is the representation for the corresponding label. Recall that in this task relation graph, $y_i$ is the label for the $i$-th task, and it has two values; similarly for each edge $\langle y_i, y_j \rangle$ has four labels with a simple combination. Thus the representations for node label $y_i$ and edge label $\langle y_i, y_j \rangle$ are as follows:
\begin{equation}
\begin{aligned}
& \vh_i^{(0)}(\vx, y_i) = \vh_i^{(0)}(\vx)[y_i]\\
& \vh_{ij}^{(0)}(\vx, y_i, y_j) = \vh_{ij}^{(0)}(\vx)[y_i,y_j].
\end{aligned}
\end{equation}

With the node and edge inputs, we can then define the message-passing propagation. Notice that here we are propagating on both the node- and edge-levels.
Following the notations in~\Cref{sec:method_gnn}, for the node-level propagation we have:
\begin{equation}
\begin{aligned}
\vh_i^{(l+1)}(\vx, y_i) & = \text{MPNN}_n^{(l+1)}\Big(\vh_i^{(l)}(\vx, y_i), \{ \vh_{ij}^{(l)}(\vx, y_i, y_j) \mid \forall j, y_j \} \Big) \\
& = \text{MLP}^{(l+1)}_n  \left( \vh_i^{(l)}(\vx, y_i) + \sum_{j\in \sN(i)} \sum_{y_j=0}^{C-1} \vh_{ij}^{(l)}(\vx, y_i, y_j) \right),
\end{aligned}
\end{equation}
and for the edge-level propagation, we have:
\begin{equation}
\begin{aligned}
\vh_{ij}^{(l+1)}(\vx, y_i, y_j) 
& = \text{MPNN}_e^{(l+1)}\Big(\vh_i^{(l)}(\vx, y_i), \vh_j^{(l)}(\vx, y_j), \vh_{ij}^{(l)}(\vx, y_i, y_j)\Big)\\
& = \text{MLP}^{(l+1)}_e \left( \vh_{ij}^{(l)}(\vx, y_i, y_j) + \text{MLP}^{(l+1)}_{a}\left( \vh_i^{(l)}(\vx, y_i) + \vh_j^{(l)}(\vx, y_j)\right) \right),\\
\end{aligned}
\end{equation}
where $\text{MLP}^{(l+1)}_n(\cdot)$, $\text{MLP}^{(l+1)}_e(\cdot)$ and $\text{MLP}^{(l+1)}_a(\cdot)$ are MLP layers defined on the node-level, edge-level, and in the aggregation function from nodes to edges. All three MLP layers are mapping functions defined on $\sR^{d}\rightarrow\sR^{d}$.

%%%%%%%%%% Training Details %%%%%%%%%%
\section{Training Details} \label{sec:appendix_training_details}
To train our proposed model, we use Adam for optimization with learning rate 1e-3, and the batch size is 32 for ChEMBL-STRING 10 (due to the memory issue) and 128 for ChEMBL-STRING 50 and ChEMBL-STRING 100. We train 200 epochs on ChEMBL-STRING 10 (within 36 hours) and 500 epochs on ChEMBL-STRING 50 and ChEMBL-STRING 100 (within 2 hours). The base graph neural network for molecule representation is GIN~\cite{xu2018powerful}, and we follow the hyperparameter used in~\cite{hu2019strategies}. The base graph neural network for task embedding is GCN~\cite{kipf2016semi}. We have more detailed description of GIN and GCN in~\Cref{sec:app_gin,sec:app_gcn}. The hyperparameter tuning for all baseline methods and SGNN base models in~\Cref{sec:appendix_hyper}.

%%%%%%%%%% Training Details %%%%%%%%%%
\subsection{Hyperparameter Tuning} \label{sec:appendix_hyper}
We list the hyperparameters for baselines models and our proposed models in~\Cref{tab:hyperparameters_baselines}, including MTL, UW~\cite{kendall2018multi}, GradNorm~\cite{chen2018gradnorm}, Dynamic Weight Average (DWA)~\cite{liu2019end}, and Loss-Balanced Task Weighting (LBTW)~\cite{liu2019loss}, SGNN in~\Cref{sec:method_gnn}, SGNN-EBM in~\Cref{sec:sgnn_ebm}.

\begin{table}[htb!]
\centering
\caption{
\small
Hyperparameters for baselines and our models.
}
\begin{tabular}{l | l l}
\toprule
Model & Hyperparameters & Values \\
\midrule
\multirow{1}{*}{MTL} & Epochs & [100, 200]\\
\midrule
\multirow{1}{*}{UW} & Epochs & [100, 200]\\
\midrule
\multirow{2}{*}{GradNorm} & Epochs & [100, 200]\\
& $\alpha$ & [0.1, 0.2, 0.5] \\
\midrule
\multirow{2}{*}{DWA} & Epochs & [100, 200]\\
& T & [0.2]\\
\midrule
\multirow{2}{*}{LBTW} & Epochs & [100, 200]\\
& $\alpha$ & [0.1, 0.2, 0.5] \\
\midrule
\multirow{5}{*}{SGNN} & Epochs & [200, 500]\\
& $d$ & [50, 100]\\
& \# GIN Layer & [5]\\
& \# GCN Layer & [0, 2]\\
& \# SGNN Layer & [2] \\
\midrule
\multirow{7}{*}{SGNN-EBM} & Epochs & [200, 500]\\
& Fixed-Noise Distribution Epochs & [200, 300, 400, 1000]\\
& $d$ & [50, 100]\\
& \# GIN Layer & [5]\\
& \# GCN Layer & [0, 2]\\
& \# SGNN Layer & [0, 2, 4]\\
& $\lambda$ & [0.1, 1]\\
\bottomrule
\end{tabular}
\label{tab:hyperparameters_baselines}
\end{table}

%%%%%%%%%% NCE %%%%%%%%%%
\section{Noise Contrastsive Esitmation with Energy Tilting Term} \label{sec:appendix_energy_tilting_term}

Here we present the derivation of the training objective function of NCE learning with tilting term in~\cref{sec:method_ebm}. When applying the backbone model for noise distribution, {\ie}, $p_n=q$, and adopting the self-normalization ($Z=1$), the loss can be rewritten as:
\begin{equation}
\begin{aligned}
\hat \gL_{NCE} & =
\mathbb{E}_{\vy \sim p_n} \log \frac{p_n(\vy|\vx)}{p_n(\vy|\vx)+ p_{\phi}(\vy|\vx)} + \mathbb{E}_{\vy \sim p_{\text{data}}} \log \frac{p_\phi(\vy|\vx)}{p_n(\vy|\vx) + p_{\phi}(\vy|\vx)}\\
& = \mathbb{E}_{\vy \sim p_n} \log \frac{p_n(\vy|\vx)}{p_n(\vy|\vx)+  p_n(\vy|\vx) \exp(-E_\phi(\vx, \vy)) } + \mathbb{E}_{\vy \sim p_{\text{data}}} \log \frac{p_n(\vy|\vx)  \exp(-E_\phi(\vx, \vy))}{p_n(\vy|\vx) + p_n(\vy|\vx) \exp(-E_\phi(\vx, \vy))}\\
& = \mathbb{E}_{\vy \sim p_n} \log \frac{1}{1 + \exp(-E_\phi(\vx, \vy))} + \mathbb{E}_{\vy \sim p_{\text{data}}} \log \frac{1}{1 + \exp(E_\phi(\vx, \vy))}.
\end{aligned}
\end{equation}

For more detailed derivations, please check~\cite{song2021train,liu2021pre}.

\end{document}